\documentclass[twocolumn,amssymb, nobibnotes, showpacs, superscriptaddress, aps, prd]{revtex4}
\usepackage{amsmath}
\usepackage{epsfig}

\begin{document}
\title{Reply to Comment by Wolfgang Ketterle on ``Electromagnetic Wave Dynamics in Matter-Wave Superradiant Scattering" (see arXiv:1010.3915)}
\author{L. Deng}
\affiliation{Physics Laboratory, National Institute of Standards \& Technology, Gaithersburg, Maryland 20899 USA}
\author{E.W. Hagley}
\affiliation{Physics Laboratory, National Institute of Standards \& Technology, Gaithersburg, Maryland 20899 USA}
\author{M. G. Payne}
\affiliation{Physics Laboratory, National Institute of Standards \& Technology, Gaithersburg, Maryland 20899 USA}
\date{\today}

   

\maketitle

{The comment by Wolfgang Ketterle \cite{ref1} purports to present a viable model of superradiance in condensates. However, Ref. \cite{ref1} is not able to explain the red/blue pump detuning asymmetry that was first observed recently by us \cite{ref11}. It is clear from our original paper \cite{ref2} that the rate-equation-based theories of Ref. \cite{ref1} are incomplete and only model the final growth stage of the process when a red-detuned pump is used \cite{ref11}. Our theoretical framework \cite{ref2}, on the other hand, also treats the initial growth stage of superradiance and is therefore also capable of explaining the genesis of the red/blue detuning asymmetry \cite{ref11}. This is the key message of our response, which we frame in terms of reference to the specific points raised in Ref. \cite{ref1}.

{\it 1. Adiabaticity. \;} Reference \cite{ref1}'s claim about our assessment of the adiabaticity criteria is incorrect. We simply questioned whether first-order adiabatic elimination used in Refs. \cite{ref4,ref5,ref8} is adequate for detunings $< 10^{10}$ Hz when a fast field relaxation rate of $10^{12}$ Hz is assumed and the time derivative of the internally-generated field cannot be neglected. Clearly, our slow propagation theory does not have this apparent ``inconsistency". However, this has no impact whatsoever on our theory.

{\it 2. Phase matching. \;} This is not new. We have already acknowledged in Ref. \cite{ref11} that the static phase matching condition derived in Ref. \cite{ref2} is not accurate \cite{phase}. This too has no impact on the correctness of our theory.

{\it 3. Slow velocity.\;} We do not agree that slow propagation for {\it superradiance} 
has been shown by Refs. \cite{ref4,ref5}, which discuss a pump-probe Bragg experiment. We note that Ref. \cite{ref10} explicitly stated that the optical fields traveled at the speed of light in vacuum and did not affect scattering at later times. To date, no report prior to our work has contradicted that statement (see Ref. \cite{ref8}). Also Ref. \cite{ref5} clearly indicates that the authors believe these are separate physical processes. We too believe that a system with a growing spontaneously-generated field (superradiant regime \cite{ref11}) behaves very differently from a system interacting with a unidirectional probe laser field (Bragg regime \cite{ref5}). In fact, it has been shown \cite{ref7} that an externally-supplied seed laser (probe in Ref. \cite{ref5}) can easily suppress superradiance, resulting in Bragg diffraction. It is therefore not logically consistent for Ref. \cite{ref1} to both claim that Ref. \cite{ref2} does not model superradiance and that Ref. \cite{ref5} (simple Bragg diffraction) proves the slow propagation of light in superradiance. Finally, Refs. \cite{ref4,ref5} only give a generic textbook expression of group velocity $v=c(n+\omega dn/d\omega)^{-1}$ to explain the slow wave in the Bragg experiment because these theories are incapable of providing more detail. 

{\it 4. Growth mechanism.\;} 
It is incorrect to say that Ref. \cite{ref2}, where both detailed atomic response and coherent propagation are treated on equal footings, does not have a feed-back mechanism that can lead to superradiance. The feed-back mechanism and the optical density \cite{linearloss} referred to in Ref. \cite{ref1} are all properly included in Eqs. (2, 3) of Ref. \cite{ref2}, where at late times in the scattering process the atomic polarization can be viewed as the matter-wave grating described in Ref. \cite{ref10}. In fact, the lack of scattering with blue detunings does indeed result from suppression of early-stage growth far from the ``high gain" (grating) superradiant threshold. The apparent confusion may come from our use of the phrase ``{\it superradiantly generated field}" immediately after Eq. (8) in Ref. \cite{ref2}. It should have been ``{\it internally generated field}", since this is precisely what Eq. (7) is about.

{\it 5. Theory. \;} The theory described in Refs. \cite{ref1, ref4, ref5, ref10} has indeed become widely accepted, but it fails to account for recent experimental results \cite{ref11}. This theory focuses only on the matter-wave grating's impact on scattering, and ignores the role of the internally-generated field which is responsible for the initial growth of matter-wave coherence. Consequently, it is incapable of explaining the observed detuning asymmetry with BECs. Also, Ref. \cite{ref4} explains superradiance with a simple rate equation plus an argument of Raman time gain based on Fermi's golden rule. In fact Eq. (1) of Ref. \cite{ref5}, which Ref. \cite{ref1} relies on, is simply a standard textbook expression of a two-photon scattering cross-section that is applicable to any two-photon process, be it in a warm vapor or in a condensate. In rate-equation-based theories \cite{ref4, ref5} it is impossible to accurately predict any propagation dynamics.  Although there have been many theoretical studies of superradiance \cite{ref8}, no analytical treatment of field propagation dynamics for collective atomic recoil motion has been reported prior to our work \cite{ref2}.

{\it Conclusion.\;} Matter-wave superradiance, like optical superradiance, is relatively simple to demonstrate experimentally 
but difficult to treat theoretically. We emphasize that the widely-accepted theoretical framework of Rayleigh scattering by a matter-wave grating \cite{ref4, ref10} is fundamentally incapable of explaining the red/blue pump asymmetry, or any 
wave propagation effects. Since this framework provides the foundation for many important studies, its revision should be a scientific priority.


\begin{thebibliography}{13}
\bibitem{phase} We believe that a dynamic matching relation including the optical dipole potential (included in Ref. \cite{ref2}), and the BEC structure factor will replace the static matching condition.
\bibitem{linearloss} Our analysis is also not predicated on a classical seed field as it can be derived using an operator formalism to describe single-photon seeds from spontaneous scattering.  The inclusion of linear loss is for the completeness only and has no bearing on the conclusion of Ref. \cite{ref2}.
\end{thebibliography}
\end{document}